%% file: main.tex
\documentclass[conference]{IEEEtran}
\makeatletter
\def\@IEEENORMtitlevspace{-21pt} %
\makeatother
\IEEEoverridecommandlockouts
\usepackage{cite}
\usepackage{amsmath,amssymb,amsfonts}
\usepackage{algorithmic}
\usepackage{graphicx}
\usepackage{textcomp}
\usepackage{xcolor}
\usepackage{multirow}
\usepackage{booktabs}
\usepackage{graphicx}
\usepackage{subcaption}
\usepackage{graphicx}
\usepackage{comment}
\usepackage{amsmath}
\usepackage{hyperref}
\usepackage{booktabs}
\usepackage{multirow}
\usepackage{algorithmic}
\usepackage{graphicx}
\usepackage{textcomp}
\usepackage{xcolor}
\usepackage{siunitx}
\usepackage{makecell}
\usepackage{caption}
\usepackage{mathrsfs}
\usepackage{bm}      
\usepackage{subcaption}  %
\usepackage{bbding}
\usepackage{cite}
\usepackage{tipa}

\newcommand{\tabsubref}[2]{Table~\ref{#1}-\subref{#2}}
\newcommand{\figsubref}[2]{Fig.~\ref{#1}(\subref{#2})}
\newcommand{\tabindent}{\hspace{1em}}
\def\BibTeX{{\rm B\kern-.05em{\sc i\kern-.025em b}\kern-.08em
    T\kern-.1667em\lower.7ex\hbox{E}\kern-.125emX}}
\begin{document}

\title{Subphonetic Acoustic Modeling via Optimal Transport for Pronunciation Assessment\vspace{-10pt}}

\author{
\IEEEauthorblockN{\small
Haopeng Geng\textsuperscript{1,*},
Jiun-Ting Li\textsuperscript{2,*},
Daisuke Saito\textsuperscript{1},
Nobuaki Minematsu\textsuperscript{1}
}
\IEEEauthorblockA{\footnotesize
\textsuperscript{1}Graduate School of Engineering, The University of Tokyo, Japan \\
\textsuperscript{2}Advanced Technology Laboratory, Chunghwa Telecom Co., Ltd., Taiwan \\
\{kevingenghaopeng, dsk\_saito, mine\}@gavo.u-tokyo.ac.jp, jtli@cht.com.tw
}
\thanks{\textsuperscript{*}Equal contribution.}
}

\maketitle

\begin{abstract}
Pronunciation assessment requires acoustic evidence that is temporally precise, diagnostically meaningful, and faithful to the learner's actual production. However, existing acoustic models often struggle to provide recognition and segmentation evidence simultaneously. CTC-based phone recognizers can predict phone sequences flexibly, but their sparse and peaky posteriors often miss phone boundaries and fine-grained pronunciation cues. In contrast, text-dependent forced aligners provide reliable temporal information when transcripts are available, but are not directly applicable to reference-free pronunciation analysis. In this work, we propose a topology-aware frame-wise acoustic model that learns dense ordered state posteriors within each phone. The key idea is to recover phone-internal state structure in a neural acoustic model by combining ordered subphonetic states with optimal temporal transport classification (OTTC). This combination encourages dense monotonic frame-level state discrimination while preserving phone recognition ability. Experiments on read, spontaneous, and L2 speech show improved segmentation over neural baselines with competitive recognition performance. Downstream evaluations further show gains in mispronunciation detection and automatic pronunciation assessment. Probing analysis suggests that the learned states capture phoneme-dependent acoustic structure rather than arbitrary frame-level distributions.

\end{abstract}

\begin{IEEEkeywords}
pronunciation assessment, forced alignment, phone segmentation, acoustic modeling, connectionist temporal classification, optimal transport
\end{IEEEkeywords}

\section{Introduction}
\label{sec:introduction}
Pronunciation assessment (PA) aims to provide learners with fine-grained feedback on how their speech deviates from target-language production \cite{zechner2025challenges}. This requires an acoustic model to answer not only which phone was produced, but also where it was realized and how its internal acoustic trajectory departs from the target. Such temporally precise and phone-internal evidence supports goodness-of-pronunciation (GOP) scoring \cite{WITT200095}, mispronunciation detection and diagnosis (MDD) \cite{7752846, leung2019cnn}, and automatic pronunciation assessment (APA) \cite{GOPT, cao24b_interspeech}, which are crucial for computer aided pronunciation training (CAPT) tasks.

This requirement exposes a tension between recognition and segmentation. Forced alignment (FA) models can produce reliable phone boundaries when a transcript is provided \cite{Rabiner89-ATO, young2002htk, mcauliffe2017montreal}, but they do not provide reliable phone hypotheses in reference-free settings. In contrast, neural phone recognition (PR) models can predict phone sequences without a fixed transcript, but often discard fine-grained temporal dynamics and phone boundaries.

Connectionist temporal classification (CTC) based PR is a representative example of this trade-off \cite{graves2006connectionist}. 
By marginalizing over possible frame-to-label alignments, CTC enables phone-sequence prediction without explicit frame-level supervision \cite{graves2006connectionist}. However, 
although effective under general PR tasks, CTC has been reported to be suboptimal in pronunciation-assessment-related tasks such as MDD \cite{geng2026beyond, lin2022phoneme}. This is because subphonetic cues, which often reveal non-canonical pronunciation patterns, are largely neglected by the CTC criterion. Moreover, the blank-dominated and peaky posteriors of CTC \cite{zeyer2021does, tian2022bayes, yao2023delay} make it difficult to predict precise phone boundaries as well.

To address these problems, we introduce a topology-aware frame-wise acoustic model trained with optimal temporal transport classification (OTTC)~\cite{kaloga2025ottc}. The model expands each phone into ordered subphonetic states and learns dense monotonic frame-to-state assignments, bringing HMM-style phone-internal topology into neural pronunciation modeling. Rather than optimizing PR accuracy alone, the model is also designed to provide temporally dense acoustic evidence for PA. The main contributions are as follows:
\begin{itemize}
    \item We propose a topology-aware frame-wise acoustic model that introduces temporally ordered subphonetic states for PA.
    \item We combine the proposed topology with OTTC to obtain dense monotonic alignments while retaining competitive PR performance.
    \item We evaluate the model on read speech, spontaneous speech, and L2 speech, covering both text-dependent forced alignment (TDFA) and text-independent (TI) segmentation/recognition.
    \item We further validate the learned acoustic evidence on L2 PA tasks, and analyze how the learned topology states represent phoneme-dependent characteristics.
\end{itemize}
\section{Research Background}
\input{figs/Model_only}
In this section, we revisit conventional forced alignment methods and neural phonetic aligners, and then explain the motivation behind the proposed framework.

\subsection{Recognition--segmentation Trade-off}
\label{sec:rec-seg_trade_off}
As discussed in Sec.~\ref{sec:introduction}, current acoustic modeling approaches often struggle to provide precise temporal localization and flexible phone recognition at the same time.
For example, classical forced alignment systems based on hidden Markov models (HMMs) recover a dense monotonic frame-to-phone path from a given transcript and can provide stable phone boundaries \cite{Rabiner89-ATO, young2002htk, mcauliffe2017montreal}. Their strength comes from using the reference sequence as a strong constraint, where each frame is assigned to a corresponding HMM state. This constraint supports accurate boundary estimation, but also makes decoding less flexible in reference-free settings. In contrast, CTC-based phone recognizers can infer phone sequences under weaker constraints, but their repeat-and-collapse objective encourages sparse non-blank emissions, making them unreliable for boundary estimation~\cite{zeyer2021does}. 
Several CTC variants have attempted to mitigate this limitation. For example, Huang et al. introduced label priors to dynamically suppress blank probabilities~\cite{LPCTC}. However, their improvements in segmentation accuracy remain limited.

Recent neural-network-based alignment models attempt to move beyond this trade-off by recovering better alignment behavior while retaining recognition ability. Charsiu~\cite{9746112} is a representative example: it uses self-supervised speech representations (SSL), forward-sum loss\cite{shih2021rad}, curriculum learning, and frame-wise cross-entropy supervision to improve phone-to-audio alignment \cite{9746112}. This line of work shows that neural acoustic models can balance recognition and segmentation more effectively. Meanwhile, as neural networks already provide better phone-level recognition performance, less attention has been paid to modeling acoustic structure below the phoneme level.

\subsection{Subphonetic Acoustic Modeling}
Another reason HMM-based models remain effective for segmentation is their use of multiple ordered states for each acoustic unit. In HMM-based ASR, a phone is commonly represented by several left-to-right (LTR) Markov states, often using a three-state topology. This structure allows the model to represent coarse phone-internal structures such as onset, steady-state, and offset regions\cite{Rabiner89-ATO, young2002htk, 266459}. Fenone \cite{150306} and Senone \cite{225979}  further refined this idea by clustering context-dependent subphonetic states for recognition. 

From this perspective, standard CTC can be viewed as a much simpler HMM, where each phone is modeled as a single label state together with the blank token. Recent CTC-topology studies also restore richer state graphs to suppress the blank ratio and improve alignment behavior \cite{10832327, 10669902}. Thus, introducing subphonetic states is not a new concept. Rather, we revisit ordered subphonetic topology as a way to provide dense and diagnostic acoustic evidence for PA. To better exploit the advantages of subphonetic modeling, we introduce a topology-aware OTTC formulation in the next section.

\section{Topology-Aware Subphonetic Modeling via Optimal Temporal Transport Classification}
\label{sec:topology_ottc_method}
\input{figs/topology_only}

In this section, we explain standard CTC and extend it to ordered multi-state subphonetic topologies. We then analyze why topology alone remains insufficient under a sparse CTC criterion, and introduce OTTC as monotonic frame-level training for learning ordered subphonetic states.

\subsection{CTC and Subphonetic Topology}
\label{sec:ctc_topology}
 Given an acoustic frame sequence $\mathbf{X}=\{x_i\}_{i=1}^{n}$ and a monophone target sequence $\mathbf{Y}=(y_1,\ldots,y_m)$, CTC optimizes the marginal probability over all possible frame-level paths that collapse to $\mathbf{Y}$:
\vspace{-0.50\baselineskip}
\begin{equation}
    \mathcal{L}_{\text{CTC}} = -\log \sum_{\pi \in \mathcal{B}^{-1}(\mathbf{Y})} \prod_{i=1}^{n} p(\pi_i \mid x_i),
    \label{eq:ctc_loss}
\end{equation}
where $\mathcal{B}^{-1}(\mathbf{Y})$ denotes the set of all paths that collapse to $\mathbf{Y}$.

Recent work has also explored CTC-based architectures by changing the topology of the output units \cite{10832327, 10669902}. Following the topology notation in~\cite{10832327} and the illustration in Fig.~\ref{fig:topology_only}, we denote the topology of an acoustic unit as S$x$T$y$. S$x$ denotes the number of ordered states used to represent the unit, while T$y$ specifies the minimum number of states that any valid traversal must consume. The ${\star}$ symbols add self-loops to selected states, allowing those states to occupy additional consecutive frames without changing the minimum traversal length. In this view, S1T1 corresponds to standard one-state CTC phone modeling, and $\text{S3T2}{\star\star}$ means that each phone can be represented by three ordered states with a minimum two-state traversal during decoding.  Let $\mathcal{P}$ denote the phone inventory. For each phone $y\in\mathcal{P}$, a topology expansion function maps it into ordered subphonetic states:
\vspace{-0.50\baselineskip}
\begin{equation}
    \tau(y)=\left(y^{(1)},y^{(2)},\ldots,y^{(K_y)}\right),
    \label{eq:topology_expand}
\end{equation}
where $K_y$ is determined by the chosen topology. The monophone sequence $\mathbf{Y}$ is then expanded into a subphonetic state sequence:
\vspace{-0.50\baselineskip}
\begin{equation}
    \mathbf{Z}=\tau(\mathbf{Y})
    =\tau(y_1)\oplus\tau(y_2)\oplus\cdots\oplus\tau(y_m),
    \label{eq:expanded_target}
\end{equation}
where $\oplus$ denotes sequence concatenation. 

\subsection{Limitations of Topology under CTC Training}
\label{sec:ctc_topology_limit}
\input{figs/CTC_OTTC}

Although topology states provide a richer label space, they do not by themselves guarantee reliable state transitions. Multi-state CTC topologies can reduce the blank ratio, but neighboring topology states may still collapse in practice~\cite{10669902}. As illustrated in Fig.~\ref{fig:CTC_OTTC_compare}, even when a phone is expanded into S3 states, its posterior may cover only part of the ordered state sequence. Thus, although topology specifies the possible LTR state order, the CTC objective can still satisfy sequence prediction without assigning dense evidence to every state. This makes the resulting posteriors still unreliable for boundary estimation and fine-grained pronunciation analysis.

\subsection{OTTC as Dense Monotonic State Transport}
\label{sec:ottc}

To address this limitation, we introduce OTTC~\cite{kaloga2025ottc}. OTTC formulates alignment as a one-dimensional mass transport problem. Instead of marginalizing over sparse CTC paths, it transports frame-level source mass to the expanded topology-state sequence $\mathbf{Z}$ and optimizes the posterior under this dense monotonic transport plan. This encourages the ordered states to receive frame-wise acoustic evidence rather than only isolated recognition peaks. Let $\mathbf{Z}=(z_1,\ldots,z_M)$ denote the expanded subphonetic state sequence derived from $\mathbf{Y}$. The frame weights $\alpha \in \Delta^n$ are predicted from the acoustic sequence by a neural network $W$:
\begin{equation}
    \alpha[\mathbf{X}, W] = \mathrm{softmax}(W(x_1), \ldots, W(x_n))^\top,
    \label{eq:alpha}
\end{equation}
and the target weights $\beta \in \Delta^M$ are typically fixed as a uniform distribution over target labels. The alignment plan $\gamma(\alpha,\beta)$ is obtained by solving
\vspace{-0.5\baselineskip}
\begin{equation}
    \gamma(\alpha, \beta) = \arg\min_{\gamma \in \Gamma_{\alpha,\beta}} \sum_{i=1}^{n} \sum_{j=1}^{M} \gamma_{i,j} \, \lVert i - j \rVert_2^2,
    \label{eq:ot_plan}
\end{equation}
where $\Gamma_{\alpha,\beta} = \left\{ \gamma \in \mathbb{R}_+^{n \times M} \mid \gamma \mathbf{1}_M = \alpha, \, \gamma^\top \mathbf{1}_n = \beta \right\}$ denotes valid couplings preserving the marginal distributions. Based on the \textit{sequence optimal transport distance} framework~\cite{kaloga2025ottc}, the OTTC loss becomes:
\vspace{-0.50\baselineskip}
\begin{equation}
    \mathcal{L}_{\text{OTTC}} = -\sum_{i=1}^{n} \sum_{j=1}^{M} \gamma_{i,j}(\alpha, \beta) \cdot \log  p(z_j \mid x_i),
    \label{eq:ottc_loss}
\end{equation}
where $p(z_j \mid x_i)$ is the posterior probability of the $j$-th topology state. As shown in Fig.~\ref{fig:CTC_OTTC_compare}, this transport process assigns probability mass to all topological states and produces a more complete subphonetic trace.

\subsection{Topology-aware Decoding}
\label{sec:topology_ottc}

At inference time, the topology is applied as a finite-state decoding constraint:
\vspace{-0.50\baselineskip}
\begin{equation}
\begin{aligned}
    \hat{\mathbf{s}} =
    \arg\max_{\mathbf{s}\in\mathcal{A}_{\tau}}
    \sum_{i=1}^{n}\log p(s_i\mid x_i),\\
    \hat{\mathbf{Z}}=\mathrm{collapse}_{\tau}(\hat{\mathbf{s}}),
    \qquad
    \hat{\mathbf{Y}}=\kappa_{\tau}(\hat{\mathbf{Z}}).
\end{aligned}
    \label{eq:topology_decode}
\end{equation}
where $\mathcal{A}_{\tau}$ is the set of valid frame-level state paths under the chosen topology. The decoded path is first collapsed into a subphonetic-state sequence $\hat{\mathbf{Z}}$, and $\kappa_{\tau}(\cdot)$ then maps this valid topology-state sequence back to phone labels. This distinction is important: an isolated substate is not sufficient to emit a phone. The decoded path must follow the state-transition rules illustrated in Fig.~\ref{fig:topology_only}.

In TDFA, the reference monophone sequence $\mathbf{Y}$ is expanded into a topology-constrained graph, and a finite-state aligner\footnote{Implemented with K2-FSA: \url{https://github.com/k2-fsa/k2}.} recovers the best path from the posteriors \cite{povey2021k2}. In the TI setting, no reference graph is supplied; we decode under the topology constraints and collapse only valid paths back to phone labels and boundaries.

\section{Experimental Evaluation}
We evaluate the proposed acoustic model from three perspectives: segmentation accuracy, acoustic fidelity, and downstream PA performance. 

\subsection{Datasets and Experimental Setup}
\label{sec:exp_setup}

\subsubsection{Datasets}
\input{datasets}

\input{performance_summary}
\input{ablation}
As summarized in Table~\ref{tab:datasets_l2}, LibriSpeech \cite{librispeech} is the only corpus used for training the proposed acoustic models. 
Specifically, TIMIT \cite{garofolo1993darpa}, Buckeye \cite{pitt2005buckeye}, and L2-ARCTIC \cite{zhao2018l2arctic} provide human-annotated phone-level timestamps for segmentation and forced alignment evaluation, while speechocean762 (SO762) \cite{zhang2021speechocean762} is used for pronunciation-related evaluation. All phone labels are mapped to a 39-phone CMUdict inventory. 

\subsubsection{Model Design}
As shown in Fig.~\ref{fig:Model_only}, the proposed model consists of a trainable SSL encoder followed by two prediction heads. The CTC head learns the monophone sequence $\mathbf{Y}$, while the SP head learns the expanded subphonetic-state sequence $\mathbf{Z}$ with OTTC. Specifically, we use the CTC posterior to condition the SP head through FiLM-style linear modulation \cite{3504035.3504518}. For each frame $i$, let $\mathbf{p}^{\mathrm{ctc}}_i$ denote the CTC posterior and $\mathbf{h}^{\mathrm{sp}}_i$ denote the hidden representation of the SP head. The conditioned SP representation is computed as
\vspace{-0.35\baselineskip}
\begin{equation}
    \widetilde{\mathbf{h}}^{\mathrm{sp}}_i
    =
    \left(1+g(\mathbf{p}^{\mathrm{ctc}}_i)\right)
    \odot \mathbf{h}^{\mathrm{sp}}_i
    +
    b(\mathbf{p}^{\mathrm{ctc}}_i),
\end{equation}
where $g(\cdot)$ and $b(\cdot)$ are linear projections and $\odot$ denotes element-wise multiplication. The overall training objective combines the CTC loss and the topology-aware OTTC loss:
\vspace{-0.35\baselineskip}
\begin{equation}
    \mathcal{L}
    =
    \lambda \mathcal{L}_{\mathrm{CTC}}
    +
    (1-\lambda)\mathcal{L}_{\mathrm{OTTC}},
\label{eq:criteria}
\end{equation}
where $\lambda=0.5$ in our setting. 

\subsubsection{Training Setup}

We trained two frame-rate variants of the proposed model. Ours--20ms uses the standard WavLM-large~\cite{chen2022wavlm} with output resolution at 20 ms, while Ours--10ms changes the final convolutional stride to obtain 10 ms frame-level outputs. The encoder and both prediction heads are trained jointly for up to 50 epochs with a batch size of 32 using only phone labels and the objective in Eq.~\eqref{eq:criteria}, followed by 1000 steps of frame-wise CE supervision to stabilize silence frames, as in Charsiu~\cite{9746112}. Each prediction head is a two-layer MLP with hidden size 384 and ReLU activation. Checkpoints are selected by LibriSpeech development PER. Based on Sec.~\ref{sec:ablation}, Ours--20ms uses S2T1, while Ours--10ms uses $\mathrm{S3T2}{\star\star}$ in the main experiments.

\subsection{Evaluation Metrics}
\subsubsection{Segmentation Accuracy}
For TDFA, we report time-stamp error (TSE) \cite{9688056}:
\vspace{-3mm}
\begingroup

\begin{equation}
\mathrm{TSE}
= \frac{1}{N}\sum_{i=1}^{N}
\left(
|\hat{s}_i-s_i| + |\hat{e}_i-e_i|
\right),
\label{eq:tse_metric}
\end{equation}
\endgroup
where $(s_i,e_i)$ and $(\hat{s}_i,\hat{e}_i)$ denote the reference and predicted start and end times of the $i$-th phone segment, respectively, and $N$ is the number of evaluated phone segments. For TI segmentation, the predicted phone sequence may have a different length. For these methods, we report the R-value by treating segmentation as boundary detection over timestamp sets following \cite{strgar2023phoneme, kreuk2020self}:
\begingroup
\vspace{-1.5mm}
\setlength{\jot}{2pt}
\begin{align}
P_{\mathrm{seg}} = \frac{|\mathcal{M}_{\delta}|}{|\widehat{\mathcal{T}}|}, \quad
R_{\mathrm{seg}} &= \frac{|\mathcal{M}_{\delta}|}{|\mathcal{T}|}, \quad
\mathrm{OS} = \frac{R_{\mathrm{seg}}}{P_{\mathrm{seg}}} - 1, \notag \\
r_1 = \sqrt{(1-R_{\mathrm{seg}})^2+\mathrm{OS}^{2}}, &\quad
r_2 = \frac{-\mathrm{OS}+R_{\mathrm{seg}}-1}{\sqrt{2}}, \notag \\
\mathrm{R\text{-}value} &= 1-\frac{|r_1|+|r_2|}{2}, \label{eq:rvalue_metric}
\end{align}
\endgroup
where boundary-matching tolerance $\delta$ is set to $20$ ms.

\subsubsection{Recognition and Diagnosis Accuracy}
For phone recognition, we report phone error rate (PER):
\begingroup
\begin{equation}
\mathrm{PER} = \frac{S+D+I}{N},
\label{eq:PER}
\end{equation}
\endgroup
where $S$, $D$, and $I$ are phone substitutions, deletions, and insertions, and $N$ is the number of reference phones. On L2-ARCTIC and SO762, we compute two types of PER: canonical PER measures dictionary-derived recognition, while perceived PER tests fidelity to the learner's realized pronunciation beyond canonical bias. We also report MDD F1 following the standard detection protocol \cite{li2016mispronunciation}:
\vspace{-0.3\baselineskip}
\begingroup
\begin{equation}
\begin{aligned}
P_{\mathrm{MDD}} = \frac{\mathrm{TR}}{\mathrm{TR}+\mathrm{FR}}&,\quad
R_{\mathrm{MDD}} = \frac{\mathrm{TR}}{\mathrm{TR}+\mathrm{FA}}, \\
F1_{\mathrm{MDD}} &= \frac{2P_{\mathrm{MDD}}R_{\mathrm{MDD}}}{P_{\mathrm{MDD}}+R_{\mathrm{MDD}}},
\end{aligned}
\label{eq:rec_mdd_metrics}
\end{equation}
\endgroup
where $\mathrm{TR}$, $\mathrm{FR}$, and $\mathrm{FA}$ denote true rejection (detected mispronunciation), false rejection, and false acceptance in MDD, respectively. 

\subsubsection{Pronunciation Assessment Performance}
For APA, we evaluate phone-level mean squared error (MSE) and Pearson correlation coefficient (PCC) on SO762 following \cite{GOPT}. MSE measures the prediction error of phone-level pronunciation scores, while PCC reflects the agreement between predicted scores and human annotations. Completeness at the utterance level is omitted for compactness.
\vspace{-3mm}
\subsection{Segmentation and Recognition Performance}
Table~\ref{tab:performance_summary_v4}  summarizes the main results for text-dependent forced alignment (TDFA), text-independent (TI) segmentation/recognition, and mispronunciation detection. 

For TDFA, MFA remains the strongest topline, consistent with prior forced-alignment comparisons \cite{rousso24_interspeech}. Among neural non-MFA systems, the proposed models give the best overall segmentation trade-off: Ours--10ms obtains the highest TD R-values on both TIMIT splits and the best TI R-values on Buckeye and L2-ARCTIC, while Ours--20ms gives the lowest non-MFA TSE on TIMIT test.

The recognition and MDD results show that the segmentation gain does not come from sacrificing phone-level evidence. Ours--20ms obtains the lowest PER on TIMIT dev/test, Buckeye, and both canonical/perceived SO762 labels, and remains second-best on both L2-ARCTIC label sets after ZIPA. Meanwhile, the proposed models achieve the best MDD F1 on both L2-ARCTIC and SO762. Overall, these results suggest that the proposed model improves temporal localization while preserving robust acoustic evidence for recognition and downstream pronunciation-related tasks.

\subsection{Ablation Studies}
\label{sec:ablation}

To isolate the contribution of each component in Sec.~\ref{sec:topology_ottc_method}, we trained the ablation models from scratch on L2-ARCTIC, following common MDD practice.

\subsubsection{Effect of Training Criterion}
\tabsubref{tab:ablation_topology_objective}{tab:ablation_objective} separates the effects of topology and training criterion. Under CTC, the richer $\text{S3T2}{\star\star}$ topology gives only modest segmentation gains, suggesting that topology alone does not ensure meaningful phone-internal state traversal. The main improvement comes from replacing CTC with OTTC under the same topology, which improves TD F1 from 38.61 to 62.91, reduces TSE from 78.14 ms to 40.64 ms, and raises TI R-value from 47.60 to 68.34. LPCTC\cite{LPCTC} provides an additional CTC variation as mentioned in Sec.~\ref{sec:rec-seg_trade_off}, but remains below OTTC in segmentation. Although CTC+$\text{S3T2}{\star\star}$ has the lowest PER, OTTC achieves the best MDD F1, indicating better temporal localization without losing pronunciation-relevant acoustic evidence.

\subsubsection{Effect of Subphonetic Topologies}
\tabsubref{tab:ablation_topology_objective}{tab:ablation_topology_resolution} shows the effect of subphonetic topology at different frame resolutions.

\noindent\textbf{Segmentation.}
Finer frame resolution better supports subphonetic state transitions. At 10 ms, increasing the topology resolution yields a clear boundary gain, improving the R-value by 5.03 points and reducing TSE by 8.84 ms compared with the monophone topology. At 20 ms, the maximum gain appears at S2, while S3 topologies do not provide significant additional benefits. This suggests that 20 ms frames are often too coarse to resolve three-state phone-internal transitions, whereas 10 ms frames provide more temporal detail for longer subphonetic modeling.

\noindent\textbf{Recognition and MDD.}
\tabsubref{tab:ablation_topology_objective}{tab:ablation_topology_resolution} also shows a trade-off between recognition and diagnosis. PER is generally better at 20 ms, where the model makes fewer frame-level decisions for sequence recognition. MDD follows the opposite trend: the best 10 ms topology improves MDD F1 by an absolute 0.03 over the best 20 ms setting.

\subsection{Downstream Task: Pronunciation Assessment Performance}

\input{APA_metrics}

We further examine whether improved acoustic evidence and alignment benefit downstream APA.
Table~\ref{tab:so762_comparison} reports the results on SO762~\cite{zhang2021speechocean762} using representative APA systems.
For the GOP-NN-based APA pipeline, i.e., GOPT\cite{GOPT} in our setting, the input features depend on both frame-wise acoustic confidence and phone-level segmentation accuracy~\cite{GOPNN}.
To disentangle posterior quality from alignment quality, we evaluate three GOPT-based settings:
(i) Fixed FA, where Kaldi-TDNN-F timestamps are fixed and LogP is varied across acoustic models;
(ii) Fixed LogP, where Kaldi-TDNN-F LogP is fixed and timestamps are varied across aligners; and
(iii) Full model, where each acoustic model provides both LogP and timestamps.

Replacing only one component generally hurts the GOPT pipeline, suggesting that mismatched LogP and timestamp sources disrupt GOP feature calibration.
In the full setting, however, our model outperforms Charsiu and the original GOPT at the phone and word levels, achieving the best phone-level MSE/PCC and improving word-level Stress PCC to 0.382.
This confirms that finer temporal granularity and refined posterior modeling improve localized pronunciation feedback.
At the utterance level, the GOPT-based pipeline still lags behind the Kaldi TDNN-F baseline.
We attribute this to the limited long-range information provided by GOP-style LogP features, which mainly capture local phoneme-level acoustic confidence but are less suited to modeling rhythm, speaking rate, pauses, and prosodic fluency.
This is consistent with prior work reporting a trade-off between local and global pronunciation scoring~\cite{li2025multi}.

The HierCB+ConPCO\cite{conPCO} experiments examine the effect of segmentation accuracy when contextual features are also modeled. Unlike the GOPT-based setting, which uses only GOP-NN features, HierCB+ConPCO combines GOP-NN features with segmented SSL representations. Here, GOP-NN features provide local pronunciation evidence, while SSL features add contextual and prosodic cues for holistic scoring. As shown in Table~\ref{tab:so762_comparison}, the utterance-level gains obtained with our timestamps suggest that more accurate boundaries help the hierarchical scorer better exploit SSL representations for modeling fluency, prosody, and overall pronunciation quality.

\vspace{-2mm}
\section{Subphonetic Acoustic Modeling and Phoneme-Dependent Characteristics}
 \input{figs/ctc_key_state_illu}
To demonstrate that the proposed SP states encode acoustic structure rather than arbitrary frame-level partitions, we analyze the Ours--10ms model from two complementary perspectives. First, we compute the S3 state occupancy for each phone and then aggregate phones by manner group, with representative phones summarized in \figsubref{fig:ctc_key_state_stat}{tab:manner_groups}. 
\figsubref{fig:ctc_key_state_stat}{fig:ctc_peak_state_manner} shows that vowels, fricatives, and liquids/glides have relatively balanced state distributions, whereas stops/affricates and nasals behave differently: stops/affricates place a 57\% share on S2, while nasals are dominated by S0. This suggests that the learned topology is not a uniform three-way partition across phonemes, but reflects manner-dependent acoustic structure.

Second, we use the CTC conditioning head shown in Fig.~\ref{fig:Model_only} as a probe. Specifically, across five random seeds, we train the CTC conditioning branch to a comparable recognition level (LibriSpeech dev PER below 4.5\%) and locate the frame with the maximum non-blank CTC posterior within each reference phone interval. The relative position is normalized to $[0,1]$, where 0 and 1 denote the first and last frames of the phone interval. As illustrated in Fig.~\ref{fig:ctc_key_state_example}, this peak provides an acoustic landmark for asking where the most discriminative evidence falls inside the ordered S3 topology.

The CTC probe reveals a clear vowel--consonant contrast. For vowels, the average CTC peak position occurs later in the phone interval, around 0.62. In contrast, consonants show earlier peak positions. Notably, stops/affricates have the earliest peak center, around 0.19, which is consistent with their strong onset or release cues; phones such as /b/ and /k/ concentrate much of their discriminative evidence near the beginning of the phone interval. These observations align with previous research
\cite{wu2007phoneme, 1660029} and acoustic phonetics\cite{wayland2018phonetics}: vowels are typically characterized by relatively stable central steady-state regions, whereas consonants often rely on short transient cues such as closure, release, or onset transitions.

The learned topology therefore does more than improve metrics: it captures phoneme-dependent acoustic structure below the phone level. This also suggests a possible diagnostic use beyond the present experiments. For example, vowel quality could be measured around nucleus-like central or later states, while consonantal errors could be examined around onset-, release-, or transition-related states, yielding a content-aware GOP-like feature for PA.

\vspace{-2mm}
\section{Conclusion and Future Work}
\input{figs/ctc_key_state_stat}

This paper presents a subphonetic acoustic modeling framework that bridges HMM-inspired ordered state modeling and end-to-end alignment learning through OTTC. Experiments across read, spontaneous, and L2 speech demonstrate improvements in forced alignment and text-independent segmentation while maintaining competitive recognition performance, with further gains in downstream APA. Ablation studies and probing show that the learned states capture meaningful phone-internal acoustic structure. In future work, we will further investigate how these subphonetic posteriors can be used for more interpretable pronunciation measurements.

\section{Ethics Statement}
We hereby acknowledge that all co-authors of this work comply with the provided code of ethics and honor the code of conduct. Our experimental corpus, Librispeech, TIMIT, Buckeye, L2-ARCTIC, and speechocean762, are widely
used and publicly available. We think there are no
potential risks for this work.

\section{Generative AI Use Disclosure}
Generative AI tools (Claude) were used for partial code writing to conduct experiments.

\bibliographystyle{IEEEtran}
\bibliography{mybib}

\end{document}

%% file: figs/Model_only.tex
\begin{figure}[t]
    \centering
    \includegraphics[width=0.75\linewidth]{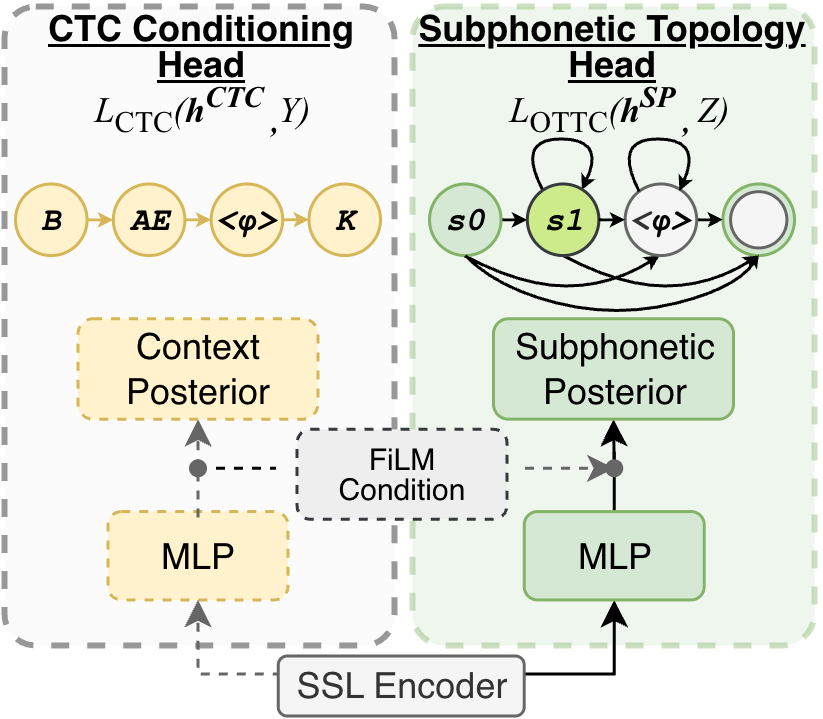}
    \caption{Illustration of proposed subphonetic model.}
    \label{fig:Model_only}
    \vspace{-2mm}
\end{figure}

%% file: figs/topology_only.tex
\begin{figure}[t]
    \centering
    \includegraphics[width=\linewidth]{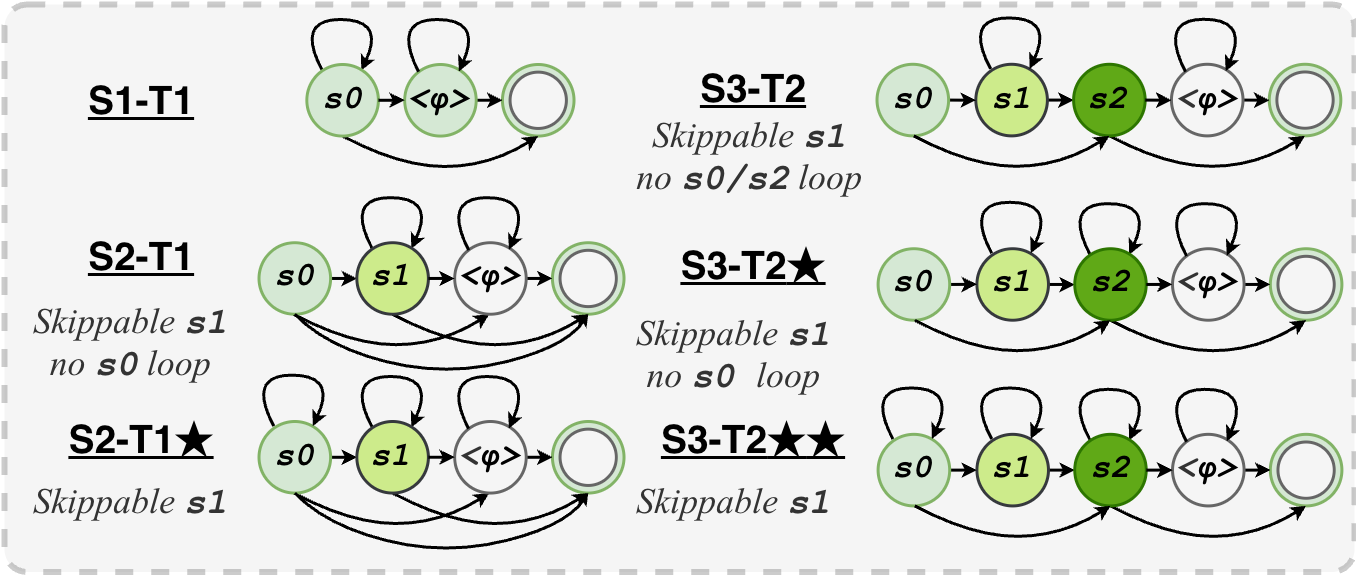}
    \caption{Illustration of subphonetic topologies.}
    \label{fig:topology_only}
\end{figure}

%% file: figs/CTC_OTTC.tex
\begin{figure}[t]
    \centering
    \setlength{\tabcolsep}{2.5pt}
    \includegraphics[width=\linewidth]{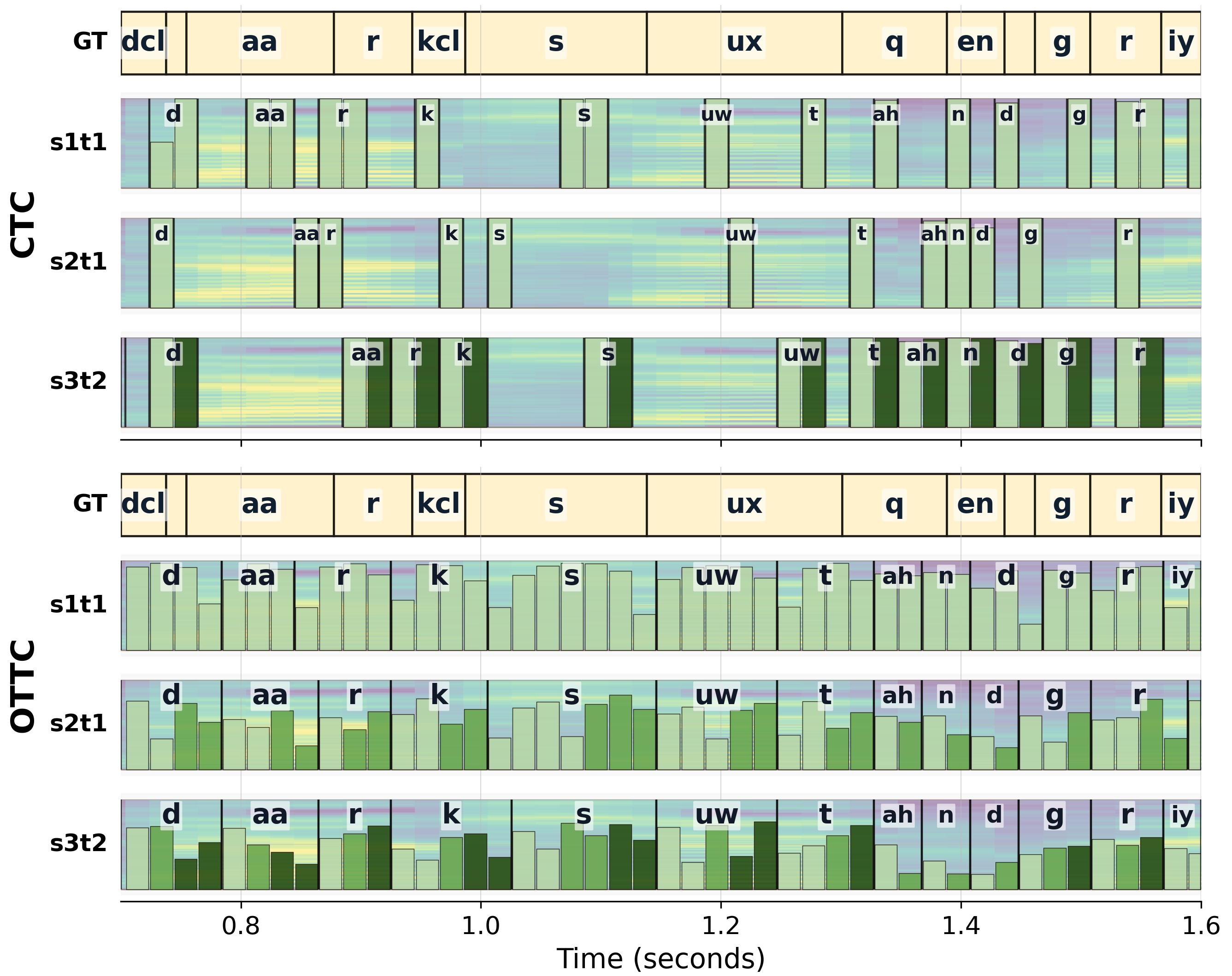}
    \vspace{-6mm}
    \caption{Comparison of frame-level probability distributions across training criteria and topologies. Different topology states are shown in different colors. CTC produces sparse spikes, while OTTC produces a dense and complete subphonetic trace.}
    \label{fig:CTC_OTTC_compare}
    \vspace{-1mm}
\end{figure}

%% file: datasets.tex
\begin{table}[t]
\centering
\caption{Datasets for training, validation, and evaluation.}
\vspace{-2mm}
\label{tab:datasets_l2}
\scriptsize
\renewcommand{\arraystretch}{1.03}
\setlength{\tabcolsep}{1.15pt}
\resizebox{\columnwidth}{!}{
\begin{tabular}{@{}l|lccccc@{}}
\toprule
\textbf{Use}
& \textbf{Dataset} 
& \textbf{Split} 
& \textbf{Dur.} 
& \textbf{Utt.} 
& \textbf{Spk.} 
& \textbf{Phone Time.} \\
\midrule

Train/Val.
& LibriSpeech\cite{librispeech}
& Train+Dev
& 971.6h
& 286.8k
& 2411
& MFA-based\footnotemark \\

\midrule

\multirow{4}{*}{Eval.}
& TIMIT\cite{garofolo1993darpa} 
& Dev/Test 
& 0.50h 
& 400/192 
& 50/24 
& Expert \\

& Buckeye\cite{pitt2005buckeye} 
& Tr/Dev/Test 
& 18.51h 
& 7946 
& 40 
& Expert \\

& L2-ARCTIC\cite{zhao2018l2arctic} 
& Test 
& 3.66h 
& 900 
& 6 
& Expert \\

& SO762\cite{zhang2021speechocean762} 
& Test 
& 5.58h 
& 2500 
& 125 
& -- \\

\bottomrule
\end{tabular}
}
\vspace{-2mm}
\end{table}
\footnotetext{Source:  \url{https://huggingface.co/datasets/gilkeyio/librispeech-alignments}.}

%% file: performance_summary.tex
\begin{table*}[t]
\centering
\caption{Baselines and proposed acoustic model's segmentation and recognition performance. Cano. and Perc. denote the canonical and perceived  phone labels used in L2 corpora, respectively.}
\label{tab:performance_summary_v4}
\vspace{-2mm}
\scriptsize
\setlength{\tabcolsep}{1.8pt}
\renewcommand{\arraystretch}{0.88}
\resizebox{\textwidth}{!}{
\begin{tabular}{lccccccc|ccccccccc}
\toprule
\multirow{4}{*}{Model}
& \multicolumn{7}{c|}{\textbf{Segmentation}}
& \multicolumn{9}{c}{\textbf{Recognition / Acoustic Fidelity}} \\
\cmidrule(lr){2-8}
\cmidrule(lr){9-17}
& \multicolumn{4}{c}{Text-dependent FA}
& \multicolumn{3}{c|}{Text-independent}
& \multicolumn{7}{c}{PER$\downarrow$}
& \multicolumn{2}{c}{MDD F1$\uparrow$} \\
\cmidrule(lr){2-5}
\cmidrule(lr){6-8}
\cmidrule(lr){9-15}
\cmidrule(lr){16-17}
& \multicolumn{2}{c}{TIMIT Dev}
& \multicolumn{2}{c}{TIMIT Test}
& \multicolumn{3}{c|}{R-value$\uparrow$}
& \multicolumn{2}{c}{TIMIT}
& \multirow{2}{*}{Buckeye}
& \multicolumn{2}{c}{L2-ARCTIC}
& \multicolumn{2}{c}{SO762}
& \multirow{2}{*}{L2-ARCTIC}
& \multirow{2}{*}{SO762} \\
\cmidrule(lr){2-3}
\cmidrule(lr){4-5}
\cmidrule(lr){6-8}
\cmidrule(lr){9-10}
\cmidrule(lr){12-13}
\cmidrule(lr){14-15}
& TSE$\downarrow$
& R-value$\uparrow$
& TSE$\downarrow$
& R-value$\uparrow$
& TIMIT Test
& Buckeye
& L2-ARCTIC
& Dev.
& Test
& 
& Cano.& Perc.& Cano.& Perc.& 
& \\
\midrule
  Kaldi TDNN-F$^\dagger$~\cite{povey18_interspeech}
  & 45.39 & 63.59
  & 45.41 & 63.53
  & - & -  & -
  & \underline{20.91} & 24.30 & 34.36
  & 25.05 & 27.27 & 54.73 & 54.44
  & \underline{0.474} & 0.156 \\
ZIPA$^\ddagger$~\cite{zipa}
& 79.21 & 55.16
& 78.93 & 56.51
& 55.23& 53.71 & 53.72
& 23.32 & 26.42 & 32.62 & \textbf{12.03} & \textbf{19.79} & 44.47 & 44.11 & 0.377 & 0.147 \\

Charsiu FC--20ms~\cite{9746112}
& 40.72 & \underline{76.55}
& 40.36 & 76.38
& \textbf{76.28} & 70.02 & 75.99
& 22.04 & \underline{24.17} & \underline{31.83} & 14.68 & 21.73 & \underline{43.93} & \underline{43.49} & 0.431 & 0.188 \\

Charsiu FC--10ms~\cite{9746112}
& 42.47 & 75.03
& 42.59 & 74.46
& 74.73 & 68.53 & 74.30
& 23.33 & 26.36 & 35.13 & 15.77 & 23.18 & 47.40 & 47.19 & 0.430 & 0.204 \\

Ours--20ms
& \underline{38.01} & 76.35
& \textbf{37.40} & \underline{76.46}
& 74.87 & \underline{73.82} & \underline{78.28}
& \textbf{20.79} & \textbf{23.20} & \textbf{29.74} & \underline{12.40} & \underline{20.53} & \textbf{39.77} & \textbf{39.41} & 0.424 & \textbf{0.240} \\

Ours--10ms
& \textbf{36.37} & \textbf{77.93}
& \underline{38.12} & \textbf{78.16}
& \underline{75.38} & \textbf{74.09} & \textbf{78.94}
& 23.89 & 26.89 & 36.53 & 20.81 & 27.96 & 64.76 & 64.74 & \textbf{0.486} & \underline{0.236} \\
\midrule
MFA 3.0~\cite{mcauliffe2026montreal}& 30.73 & 87.83
& 31.95 & 87.44
& -- & -- & --
& -- & -- & -- & -- & -- & -- & -- & -- & -- \\
\bottomrule
\end{tabular}
}
\renewcommand{\arraystretch}{1.0}
\vspace{0.5mm}
\begin{minipage}{\textwidth}
\scriptsize
$^\dagger$Kaldi TDNN-F predicts over 6000 context-dependent HMM pdf-ids rather than monophones; we therefore reconstructed phone-level decoding using a monophone-target HMM.

$^\ddagger$ZIPA originally uses a 127-symbol IPA inventory. For a fair comparison, we decoded IPA sequences and mapped them to the 39-symbol ARPAbet inventory used in this study.
\end{minipage}
\end{table*}

%% file: ablation.tex
\begin{table*}[t]
\centering
\caption{Ablation on L2-ARCTIC. All models were trained from scratch on the L2-ARCTIC training set, following common practice for MDD. Notably, the CTC conditioning head in Fig.~\ref{fig:Model_only} was not applied in this ablation.}
\label{tab:ablation_topology_objective}
\scriptsize
\vspace{-2mm}
\renewcommand{\thesubtable}{\Alph{subtable}}
\captionsetup[subtable]{font=scriptsize,skip=0pt,labelformat=parens,labelsep=space,subrefformat=parens}
\begin{subtable}[t]{0.43\textwidth}
\centering
\caption{Effect of training criterion at 20 ms.}
\label{tab:ablation_objective}
\setlength{\tabcolsep}{2.8pt}
\renewcommand{\arraystretch}{0.96}
\resizebox{\linewidth}{!}{
\begin{tabular}{llccccc}
\toprule
\multirow{2}{*}{Objective}
& \multirow{2}{*}{Topology}
& \multicolumn{2}{c}{TD FA}
& \multicolumn{1}{c}{TI seg.}
& \multicolumn{2}{c}{Recog. / MDD} \\
\cmidrule(lr){3-4}
\cmidrule(lr){5-5}
\cmidrule(lr){6-7}
&
& F1$\uparrow$
& TSE$\downarrow$
& R-val$\uparrow$
& PER$\downarrow$
& MDD F1$\uparrow$ \\
\midrule
CTC
& S1T1 & 32.73 & 82.90 & 42.58 & 18.19 & 0.562 \\
CTC
& $\text{S3T2}{\star\star}$ & 38.61 & 78.14 & 47.60 & \textbf{16.68} & 0.589 \\
LPCTC \cite{LPCTC}
& S1T1 & 49.22 & 69.32 & 56.65 & 18.07 & 0.610 \\
OTTC
& S1T1 & 57.02 & 46.65 & 63.31 & 18.44 & 0.604 \\
OTTC
& $\text{S3T2}{\star\star}$ & \textbf{62.91} & \textbf{40.64} & \textbf{68.34} & 19.24 & \textbf{0.619} \\
\bottomrule
\end{tabular}
}
\end{subtable}
\hfill
\begin{subtable}[t]{0.55\textwidth}
\centering
\caption{Topology effect at different frame resolutions.}
\label{tab:ablation_topology_resolution}
\setlength{\tabcolsep}{3.0pt}
\renewcommand{\arraystretch}{0.84}
\resizebox{\linewidth}{!}{
\begin{tabular}{lcccc|cccc}
\toprule
\multirow{2}{*}{Topology}
& \multicolumn{4}{c|}{20 ms}
& \multicolumn{4}{c}{10 ms} \\
\cmidrule(lr){2-5}
\cmidrule(lr){6-9}
& TSE$\downarrow$
& R-value$\uparrow$& PER$\downarrow$
& MDD$\uparrow$
& TSE$\downarrow$
& R-value$\uparrow$& PER$\downarrow$
& MDD$\uparrow$ \\
\midrule
S1T1
& 46.65 & 63.31 & \textbf{18.44} & 0.604
& 48.09 & 66.46 & \textbf{19.87} & 0.632 \\
S2T1
& 39.99 & 68.38 & 18.79 & 0.618
& 41.49 & 69.72 & 20.89 & 0.641 \\
$\text{S2T1}{\star}$
& \textbf{39.53} & \textbf{68.72} & 18.91 & \textbf{0.623}
& 42.42 & 69.55 & 20.59 & 0.644 \\
S3T2
& 40.94 & 67.77 & 19.27 & 0.617
& 39.78 & 71.17 & 22.18 & \textbf{0.651} \\
$\text{S3T2}{\star}$
& 40.72 & 68.40 & 19.00 & 0.616
& \underline{39.59} & \underline{71.21} & 21.43 & 0.644 \\
$\text{S3T2}{\star\star}$
& 40.64 & 68.34 & 19.24 & 0.619
& \textbf{39.25} & \textbf{71.49} & 21.14 & \underline{0.645} \\
\bottomrule
\end{tabular}
}
\end{subtable}
\vspace{-2mm}
\end{table*}

%% file: APA_metrics.tex
\begin{table}[t]
\centering
\caption{
Comparison with representative APA systems GOPT~\cite{GOPT} and ConPCO~\cite{conPCO}.
Indented rows denote controlled variants that use the same scoring pipeline but replace either
the frame-level log posteriors or the phone-level timestamps. Acc., Str., Flu., Pro., and Tot. denote
accuracy, stress, fluency, prosody, and total score, respectively.
}
\label{tab:so762_comparison}
\vspace{-2mm}

\small
\setlength{\tabcolsep}{2.2pt}
\renewcommand{\arraystretch}{0.90}

\resizebox{\columnwidth}{!}{%
\begin{tabular}{lccccccccc}
\toprule
\multirow{2}{*}{Model}
& \multicolumn{2}{c}{Phone}
& \multicolumn{3}{c}{Word PCC}
& \multicolumn{4}{c}{Utt. PCC} \\
\cmidrule(lr){2-3} 
\cmidrule(lr){4-6} 
\cmidrule(lr){7-10}
& MSE $\downarrow$ 
& PCC $\uparrow$ 
& Acc. $\uparrow$ 
& Str. $\uparrow$ 
& Tot. $\uparrow$ 
& Acc. $\uparrow$ 
& Flu. $\uparrow$ 
& Pro. $\uparrow$ 
& Tot. $\uparrow$ \\
\midrule

Kaldi TDNN-F~\cite{GOPT} 
& 0.085 
& 0.612 
& 0.533 
& 0.291 
& 0.549 
& \textbf{0.714} 
& \textbf{0.753} 
& \textbf{0.760} 
& \textbf{0.742} \\

\tabindent w/ Charsiu's LogP
& 0.126 
& 0.293 
& 0.308 
& 0.094 
& 0.317 
& 0.585 
& 0.661 
& 0.660 
& 0.614 \\

\tabindent w/ Ours LogP
& 0.130 
& 0.272 
& 0.297 
& 0.040 
& 0.306 
& 0.580 
& 0.626 
& 0.635 
& 0.596 \\

\tabindent w/ Charsiu's TS
& 0.124 
& 0.300 
& 0.306 
& 0.189 
& 0.312 
& 0.517 
& 0.596 
& 0.583 
& 0.536 \\

\tabindent w/ Ours TS
& 0.123 
& 0.302 
& 0.307 
& 0.277 
& 0.310 
& 0.513 
& 0.592 
& 0.580 
& 0.531 \\

\addlinespace[3pt]

Charsiu TS\&LogP
& 0.101 
& 0.518 
& 0.464 
& 0.151 
& 0.476 
& 0.682 
& 0.705 
& 0.705 
& 0.701 \\

Ours TS\&LogP
& \textbf{0.084} 
& \textbf{0.617} 
& \textbf{0.565} 
& \textbf{0.382} 
& \textbf{0.578} 
& 0.712 
& 0.712 
& 0.717 
& 0.723 \\

\midrule

HierCB+ConPCO~\cite{conPCO} 
& \textbf{0.071} 
& \textbf{0.701} 
& \textbf{0.669} 
& \textbf{0.437} 
& \textbf{0.682} 
& 0.780 
& 0.830 
& 0.823 
& 0.803 \\

\tabindent w/ Charsiu's TS& 0.083 
& 0.635 
& 0.589 
& 0.221 
& 0.602 
& \underline{0.785} 
& \underline{0.835} 
& \underline{0.831} 
& \underline{0.812} \\

\tabindent w/ Ours TS& 0.080 
& \underline{0.658} 
& \underline{0.621} 
& 0.215 
& \underline{0.633} 
& \textbf{0.796} 
& \textbf{0.837} 
& \textbf{0.833} 
& \textbf{0.819} \\

\bottomrule
\end{tabular}
}

\renewcommand{\arraystretch}{1.0}
\vspace{-1mm}
\end{table}

%% file: figs/ctc_key_state_illu.tex
\begin{figure}[t]
            \vspace{-4pt}
            \centering
            \includegraphics[width=0.90\linewidth,trim=1.15in 0.55in 1.15in 0.55in,clip]{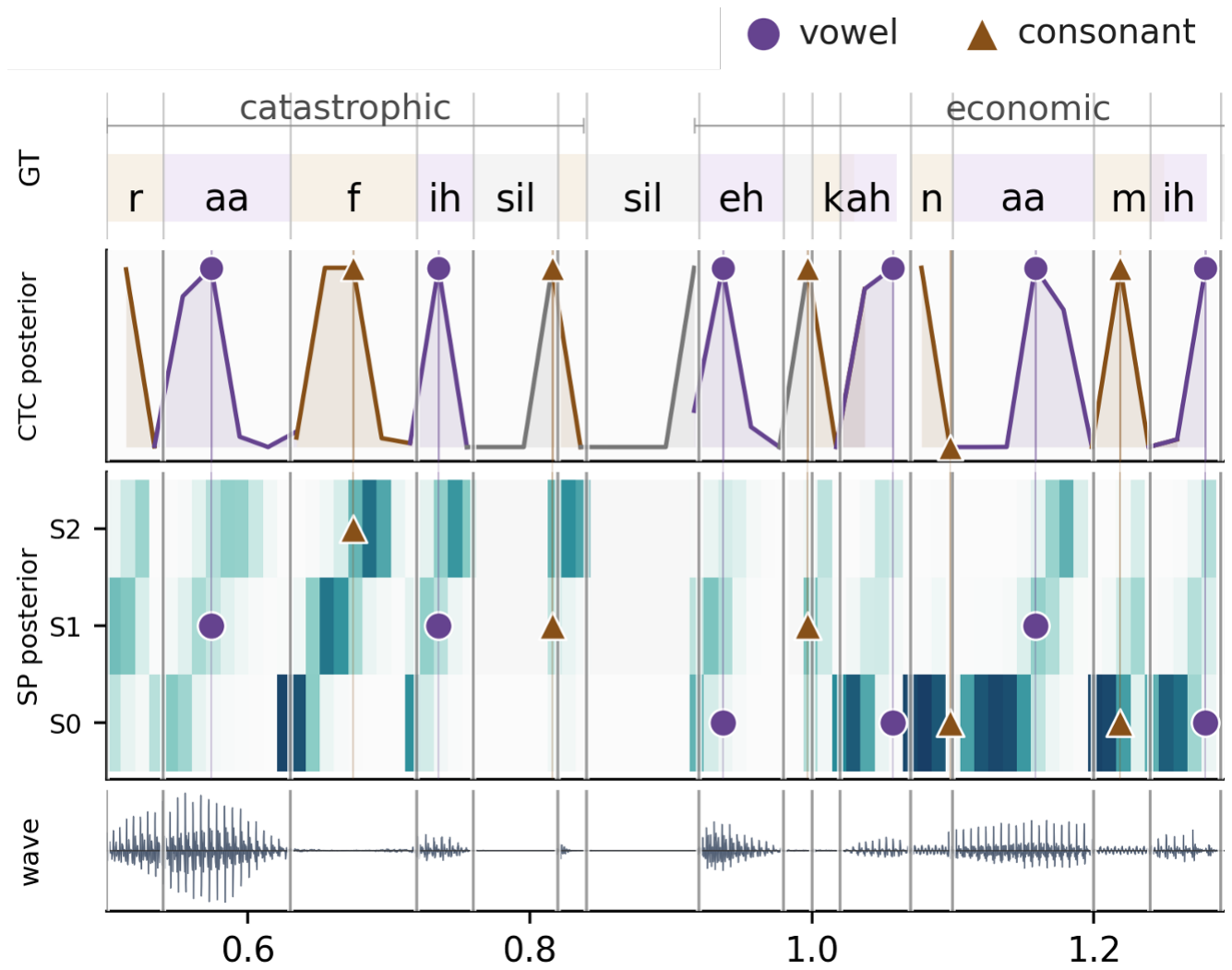}
              \caption{CTC peaks as acoustic landmarks for probing subphonetic states}
            \label{fig:ctc_key_state_example}
\vspace{-1mm}
\end{figure}

%% file: figs/ctc_key_state_stat.tex
\begin{figure}[t]
    \centering
    \begin{subfigure}[t]{.75\linewidth}
        \centering
        \vspace{0pt}
        \includegraphics[width=\linewidth,clip]{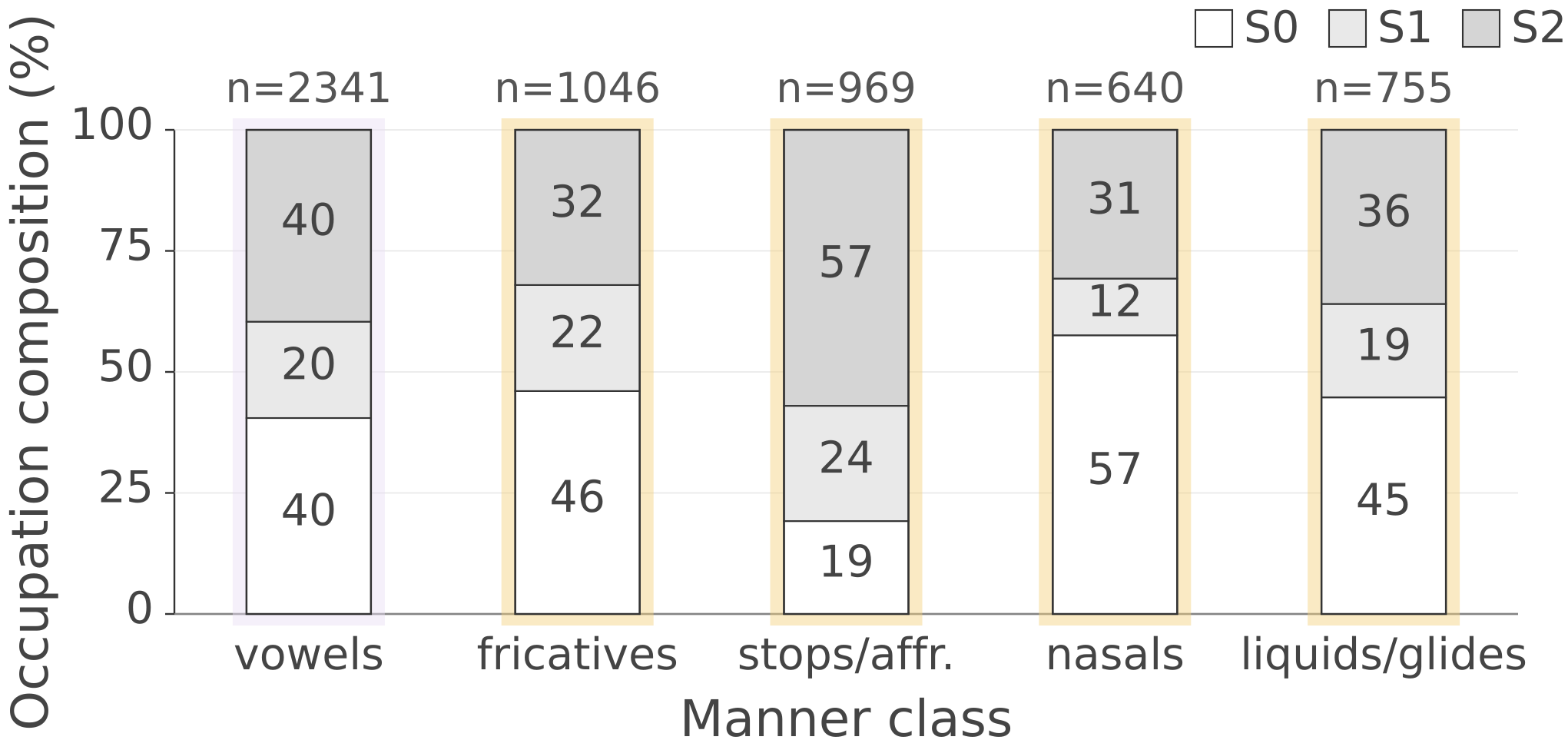}
        \vspace{-0.2\baselineskip}
        \includegraphics[width=\linewidth,clip]{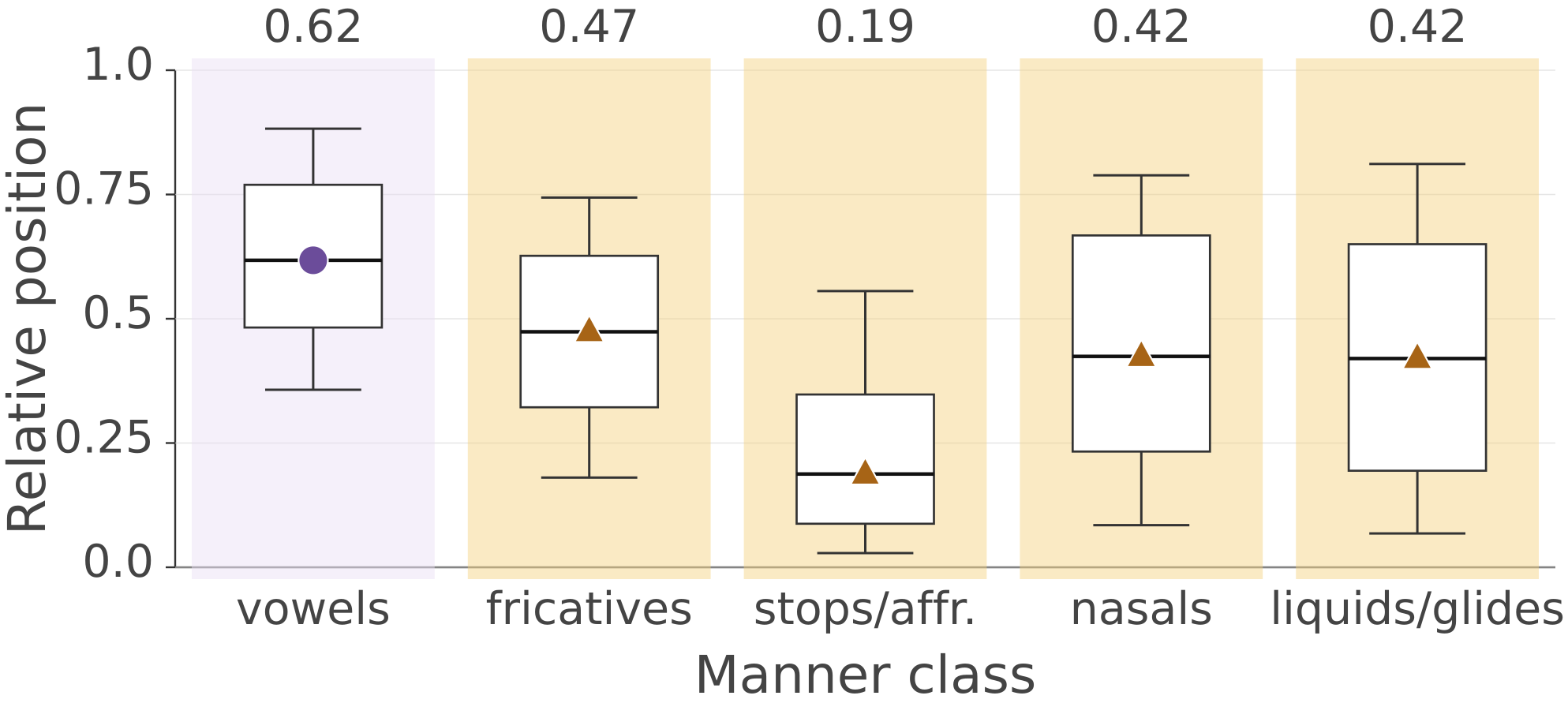}
      \captionsetup{width=1.5\textwidth}
        \caption{Articulation-Manner-level S3 occupancy and CTC peak centers.}
        \label{fig:ctc_peak_state_manner}
    \end{subfigure}
\hfill
        \begin{subfigure}[t]{\linewidth}
            \centering
            \begingroup
            \scriptsize
            \newcommand{\ph}[1]{\textsl{\texttt{#1}}}
            \setlength{\tabcolsep}{1.5pt}
            \renewcommand{\arraystretch}{0.90}
            \begin{tabular*}{\linewidth}{@{\extracolsep{\fill}}lccccc@{}}
            \toprule
            Group & Vowel. & Fricatives. & Stop/aff. & Nasals. & Liquids. \\
            \midrule
            phones
            & \ph{aa} \ph{ae} \ph{ih} \ph{uw}
            & \ph{f} \ph{s} \ph{sh} \ph{z}
            & \ph{b} \ph{d} \ph{k} \ph{ch}
            & \ph{m} \ph{n} \ph{ng}
            & \ph{l} \ph{r} \ph{w} \ph{y} \\
            \bottomrule
            \end{tabular*}
            \endgroup
            \caption{Phones grouped by the manner of articulation.}
            \label{tab:manner_groups}
        \end{subfigure}
 \caption{Subphonetic topology and CTC-probe analysis for pronunciation modeling.}
\label{fig:ctc_key_state_stat}
\end{figure}